%Paper: hep-ph/9301206
%From: LIPKINZ@taunivm.tau.ac.il
%Date: 5 Jan 93   16:25 IST

%macropackage=phyzzx
\magnification=\magstep1
\parskip 0pt
\def\({[}
\def\){]}

\def\rjustline#1{\line{\hss#1}}

\def\rarrow{\rightarrow}

\rjustline{ANL-HEP-PR-92-87}
\rjustline{WIS-92/80/Oct-PH}
\overfullrule 0pt
\centerline{\bf Implications of $\tau$ Decays into Strange Scalar
and Axial Vector Mesons}
\author{Harry J. Lipkin}
\smallskip
\centerline{Department of Nuclear Physics}
\centerline{\it Weizmann Institute of Science}
\centerline{Rehovot 76100, Israel}
\centerline{and}
\centerline{School of Physics and Astronomy}
\centerline{Raymond and Beverly Sackler Faculty of Exact Sciences
}
\centerline{\it Tel Aviv University}
\centerline{Tel Aviv, Israel}
\centerline{and}
\centerline{High Energy Physics/Physics Divisions}
\centerline{Argonne National Laboratory}
\centerline{Argonne, IL 60439}
\centerline{Revised January 3, 1993}
\vskip 0.2in
\baselineskip 12pt
\abstract
Recent reports of the observation of the decay $\tau \rarrow
\nu_{\tau} + K + \pi + \pi$ raise interesting questions.
The $u-s$ quark mass difference can induce an effective second-class
current in the strange sector and enhance decays to helicity-suppressed
scalar states.
This suggests searches for scalar strange mesons in
the $\tau \rarrow \nu_{\tau} + K + \eta' $ and
$\tau \rarrow \nu_{\tau} + K + \pi$ decay modes.
$\tau$ decays to axial vector final states may clarify
strange meson spectroscopy by analogy with the clarification
of the $a_1$ problem in the nonstrange sector.
Open questions include
mixing effects in the strange axial vector mesons and the implications of
models based on QCD for hadronization and final state interactions
following the creation of a $q \bar q$ pair by a weak vertex,
\endpage

$\tau$ decays into strange final states have recently been reported
\REF{\TPC}{TPC/Two-Gamma Collaboration, LBL preprint LBL-32377,
August 1992}
\REF{\Ronan}{M.T. Ronan, Presented at the Second Workshop on Tau Lepton
Physics, The Ohio State University, Columbus, Ohio,
USA, September 8-11, 1992. LBL preprint LBL-33038}
[\TPC,\Ronan]
and show indications for the strange axial-vector mesons in the
$\tau \rarrow \nu_{\tau} + K + \pi + \pi$ decays.
In the analogous nonstrange
decays, the three-pion final state is dominated by the $a_1$ axial vector
meson,
while the other axial vector state $b_1$ and the scalar state $a_0$ have not
been observed. The $b_1$ and $a_0$ do not couple to first-class currents and
their absence is consistent with the absence of second-class currents in the
standard model.

The transition $W \rarrow $ $ scalar$ $meson$ was shown\REF{\BergLip}{Edmond
L. Berger and Harry J. Lipkin, Phys. Lett. B 189, (1987) 226}[\BergLip]
by simple arguments to be suppressed in any
model where the $W$ decays into a quark-antiquark pair with a pointlike
coupling and the hadronization of the pair occurs subsequently.
Their treatment [\BergLip] can be applied to strange decays
$$\tau\rarrow \nu_\tau + \bar u+s\rarrow \nu_\tau +
{\rm strange~ mesons} \eqno(1) $$
by an SU(3) rotation and replacing $G-parity$ replaced by another
subgroup of flavor SU(3), which we denote by $G_{us}$, and which acts in the
$u-s$ sector in a manner analogous to conventional $G$ parity in the $u-d$
sector.

The production of a spin zero meson in the final state in the strange
decay (1) is suppressed by two symmetry selection rules. The well known
helicity selection rule forbids the coupling of a spin zero state to a
left-handed zero-mass fermion-antifermion pair.
Selection rules forbidding charged second class currents follows from
using flavor symmetry to rotate a CP selection rule for neutral
current couplings. The CP-forbiddenness of couplings of weak currents
to a spin zero state even under CP is rotated into forbidding
couplings to spin zero states with odd $GP$ and odd $G_{us}P$.

The $GP$ and $G_{us}P$ selection rules are satisfied for pseudoscalar leptonic
decays, since both the pion and kaon are even respectively under $GP$ and
$G_{us}P$. The scalar mesons are odd respectively under $GP$ and $G_{us}P$.
Their weak couplings require second class currents and are forbidden in the
flavor symmetry limit.
Symmetry breaking effects arise
at the quark level in the $W q\bar q$ vertex as a result of the quark mass
differences and also in the hadron mass spectrum and couplings. Isospin and
$G$ breaking in nonstrange $\tau$ decays are discussed in detail in ref.
[\BergLip]. To extend this treatment to the strange case we rotate
into the strange sector
their quark-antiquark states
$$\ket{\vec k,-\vec k;0;\pm } =
\left(1/\sqrt 2\right)\left(\ket{s_R(\vec k),\bar u_R(-\vec k)} \mp
\ket{s_L(\vec k),\bar u_L(-\vec k)}\right);\eqno(2)  $$
where $\ket{\bar u_L(\vec k)}$ denotes a $\bar u$ antiquark with momentum
$\vec k$ and left-handed helicity and similarly for
$\ket{\bar u_R(\vec k)}$, $\ket{s_L(\vec k)}$ and $\ket{s_R(\vec k)}$.
These states (2) are eigenstates of the transformations
$ Pe^{i\pi J_n} $ and
$G_{us}P$ with opposite eigenvalues,
$$ Pe^{i\pi J_n} \ket{\vec k,-\vec k;0;\pm} = \pm \ket{\vec k,-\vec k;0;\pm};
\eqno (3a)$$
$$ G_{us}P \ket{\vec k,-\vec k;0;\pm} = \mp \ket{\vec k,-\vec k;0;\pm}.
\eqno (3b)$$
where $P$ denotes space inversion and $J_n$ is the generator of
rotations about an axis $\vec n$ normal to the momentum $\vec k$.

The two allowed $J=0$ states for a $s \bar u$ pair are also
eigenstates of both these operations:
$$ Pe^{i\pi J_n} \ket{J^P=0^\pm} = \pm \ket{J^P=0^\pm}, \eqno(4a) $$
$$ G_{us}P \ket{J^{PG}=0^{\pm -}} = \mp
\ket{J^{PG}=0^{\pm -}}. \eqno(4b) $$
Note that the eigenvalue of $Pe^{i\pi J_n}$ for a $J=0$ state is just the
parity, since $J_n = 0$ and that the states of both parities have odd
$G_{us}$ parity. These states
have the quantum numbers of the kaon and the $K^*_0$ (1430).

If there are only zero-mass quarks and left-handed currents,
the intermediate $s \bar u$ state in the decay (1) must have spin one
in the center-of-mass system of the pair, since the spins of the left-handed
quark and the right-handed antiquark are parallel. The production of a
spin-zero state is thus forbidden for zero-mass quarks. This is just the
standard helicity selection rule which forbids the $e\nu$ decays of the pion
and kaon in the limit of zero electron mass.

For massive quarks the spin-zero state can be produced by the scalar (time)
component of the $W$. The amplitude for the production of a right-handed quark
or a left-handed antiquark by a left-handed
current is given by the Dirac equation as proportional to $(m_q/\sqrt 2 E)$
where $m_q$ is the quark mass and $E$ is the energy. Thus a zero helicity state
of a quark with momentum $\vec k$ and an antiquark with momentum $-\vec k$
can be produced in the decay of a $W$ from the time component and has the form
$$ \ket{\vec k,-\vec
k;0;W_o} = \left({{m_u}\over{\sqrt 2 E}}\right)\ket{s_L(\vec k),\bar u_L(-\vec
k)}+ \left({{m_s}\over{\sqrt 2 E}}\right)
\ket{s_R(\vec k),\bar u_R(-\vec k)} = $$
$$ =
\left({{m_s + m_u}\over{2E}}\right) \ket{\vec k,-\vec k;0;-} +
\left({{m_s - m_u}\over{2E}}\right) \ket{\vec k,-\vec k;0;+}.
\eqno(5) $$
The weak interaction chooses the particular linear
combination (5) which depends upon the quark masses.
It is not normalized; its magnitude is proportional to the amplitude for
producing the state in a weak interaction.
All terms in (5) are proportional to the ratio of some quark
mass to the energy $E$ and vanish for zero quark mass, as expected
from the helicity suppression.

Pseudoscalar meson production arises from the component
$ \ket{\vec k,-\vec k;0;-} $
proportional to
$\left({{m_s + m_u}\over{2E}}\right)$.
Scalar meson production comes from the component
$ \ket{\vec k,-\vec k;0;+} $.
This component is proportional to
$\left({{m_s - m_u}\over{2E}}\right)$
and vanishes when the two quark masses are equal. However, in contrast to
the nonstrange case, where ${{m_d - m_u}\over{2E}}$ is small, here
${{m_s - m_u}\over{2E}}$ can be of order unity and there may be
very little suppression.
Note however that theoretical models whose
predictions depend upon values of quark masses are subject to the confusion
between constituent and current quark masses. These can give very different
results for helicity suppression factors and for flavor symmetry breaking at
the weak pointlike vertex. The suppression of the forbidden scalar
decay relative to the allowed pseudoscalar decay is given by the
ratio $\left({{m_s - m_u}\over{m_s + m_u}}\right)^2$. This is $\approx 1$
when current quark masses are used and $\approx 4\%$ when constituent
quark masses are used with $m_s \approx (3/2) m_u$.
This point can be clarified by experiment.
Note that the suppression factor obtained from the initial
pointlike vertex is not changed by the hadronization process, even
though isospin symmetry and $G_{us}$-parity may be broken. Because the
strong interactions responsible for hadronization conserve angular
momentum and parity, they cannot change the eigenvalue of
$Pe^{i\pi J_n}$
and produce a forbidden state. That $G_{us}$ is not conserved is irrelevant.
The one-to-one correspondence between the eigenvalues of $G_{us}$ and the
conserved quantity $Pe^{i\pi J_n}$ also preserves the $G_{us}$ eigenvalue
in the subspace of these particular final states.

The production of the scalar meson state $K^*_0$ (1430) in $\tau$ decay
thus gives information about the SU(3) symmetry breaking at the weak
vertex and cannot be enhanced by SU(3) symmetry breaking in the hadronic wave
functions. If the symmetry breaking occurs at the current quark level
before the current quarks created by the weak vertex are somehow dressed
to become constituent quarks, the current quark masses determine the
symmetry breaking and give no suppression. This contrasts with the
suppression by a factor of 25 when constituent masses are used.
It is therefore of
interest to look for the strange scalar $K^*_0$ (1430) in $\tau$ decays.
This
state could be observed in the $K_S \pi^{\pm}$ decay mode with detectors
sensitive to charged pions.
The $\tau \rarrow \nu_{\tau} + K + \eta $ decay mode is expected to be
suppressed by interference effects between the strange and nonstrange
components of the neutral mesons which are destructive in the scalar
channel for the $\eta$ and constructive for the $\eta'$
\REF{\EtaLip}{Harry J. Lipkin, Phys. Lett. B 254, (1991) 226
and B 283, (1992) 421}[\EtaLip]. However it might be
strong enough to be observed with detectors that have good kaon identification
and can also detect the $\eta$.
The $K^{\pm} \eta'$ decay mode is at the edge of phase space, but might be
observed with detectors with good kaon identification and good $\eta'$
detection. These interference effects have been observed in hadronic charm
decays where the $K^o \eta'$ decay mode is at the edge of phase space but
nevertheless observed and is considerably stronger than the observed
$K^o \eta$ decay when the phase space factors are removed[\EtaLip].

The strange axial vector system is more complicated and more interesting
than both the pseudoscalars and the nonstrange axial vectors. There are
several polarization states, some of which have no helicity suppression,
two axial vector states with opposite eigenvalues of $G_{us}$ which can mix by
virtue of symmetry breaking, and two vector-pseudoscalar final states
$K^* \pi$ and $K \rho$ instead of the single $\rho - \pi$ state in the
nonstrange sector. Flavor symmetry requires equality of the
$K^* \pi$ and $K \rho$ branching ratios.
Symmetry breaking can occur not only at the
weak vertex as in the scalar case but also in the hadron dynamics.

In the SU(3) symmetry limit
the two axial-vector kaon states classified in the same octets as the
$a_1$ and $b_1$ and denoted
respectively by $K_A$ and $K_B$ would
be coupled
respectively to first class and second class currents  like the $a_1$ and $b_1$
and only the $K_A$ would be produced in $\tau$ decay. However,
experimental meson spectroscopy shows that these two states are very strongly
mixed, with a mixing angle of about $45^o$ expected from some
theoretical models\REF{\LIPQ}{Harry J. Lipkin, Phys. Lett. B 72 (1977)
249}[\LIPQ], giving the two states denoted as $K_1$ (1270) and $K_1$ (1400).
The $45^o$ mixing angle is indicated by the observations that the
$K_1$ (1270) decays dominantly into $K\rho$ and the $K_1 (1400)$ decays
dominantly into $K^* (892)\pi$, while the $K_A$ and $K_B$ are predicted by
SU(3)
to have equal amplitudes for both decay modes and with opposite phase. Thus
$45^o$ mixing decouples one eigenstate from $K\rho$ and the other from
$K^* (892)\pi$. More recent developments have been summarized by Suzuki
\REF{\Suzuki}{Mahiko Suzuki, UCB preprint UCB-92/32, LBL preprint LBL-32865
Sepember 1992}
[\Suzuki]
who has calculated explicitly the best current values
of mixing angles and the $\tau$ decays into these states using present data.

At the weak vertex both vector and axial vector final states have contributions
without helicity suppression from the quark-antiquark state where both quark
and antiquark have the ``correct" helicities. Since the transformations $GP$
and
$G_{us}P$ reverse helicity and change a quark into an antiquark, they change
left-handed quarks (correct helicity) into right-handed antiquarks (correct
helicity) and vice versa while changing right-handed quarks (wrong helicity)
into left-handed antiquarks (wrong helicity) and vice versa. These
transformations thus conserve the ``correctness" or ``wrongness" of helicity
while changing quarks into  antiquarks and vice versa. They therefore transform
a quark-antiquark state where both quark and antiquark have the ``correct"
helicities into itself. The phase factor introduced can be shown to give
the negative eigenvalue corresponding to a first class current and
the quantum numbers of the $\rho$ and $a_1$ mesons in the
nonstrange sector and the $K^*$(890) and the $K_A$ mesons in the strange
sector. This argument is independent of quark masses.

This state where both quark and antiquark have the ``correct" helicity
is the dominant final state in nonstrange decays, where
the helicity suppression and quark mass factors
are expected to be serious and render the second class contributions
negligible. In the strange sector these suppression factors can well be
only of order unity and lead to appreciable second class
contributions. We disregard these contributions as a first approximation
and examine the experimental implications of assuming first class
dominance.

An additional factor which must be considered in the axial vector case
is the hadronization process in which the quark-antiquark pair produced
at the weak vertex dresses itself with an additional quark-antiquark pair
and gluons to become the observed final state. There are two possible
models.

1. Resonance domination. This is the model considered[\Ronan,\Suzuki] in
the recent data analysis. The quark-antiquark pair produced at the weak
vertex dresses itself by gluon exchange to become some combination of
the states $K_A$ and $K_B$ and these decay subsequently into the observed
$K^* \pi$ or $K \rho$ final states.

2. Direct production. The quark-antiquark pair produced at the weak
vertex fragment immediately into the $K^* \pi$ or $K \rho$ final states
by the creation of an additional quark-antiquark pair. The resonances
then play a role only via the final state interactions of the two-meson
state.

These two models give different predictions which can be checked
by experiment and give useful information about how QCD makes hadrons
from quarks and gluons. In the approximation that the
first class current dominates at the weak vertex
the resonance dominance model predicts that only
the $K_A$ state is produced, with roughly equal amplitudes for
$K_1 (1270)$ and $K_1 (1400)$ and some freedom in the ratio possible
by adjusting the mixing angle.
$$ \ket{s_L,\bar u_R}
\rightarrow  \ket {K_A} \approx (1/\sqrt 2)\cdot\{\ket {K_1 (1270)}
+ \ket {K_1 (1400)}\} \eqno(6)  $$
In this case phase space favors the
$K_1$ (1270) which decays dominantly into $K\rho$. The experimental
result that only the $K_1 (1400)$ and the $K^* \pi$ decay mode is
observed [\Ronan] with an excess of 12 events at 1400 MeV
conflicts with this simple picture, and fine tuning of mixing
angles is required to obtain agreement with experiment[\Suzuki].

The direct production model predicts equal amplitudes for
fragmentation into $K^* \pi$ or $K \rho$ in the flavor-SU(3)
symmetry limit. For the case where an additional $d \bar d$ pair is
produced to give a charged nonstrange meson and a neutral strange
meson,
$$
\ket{s_L,\bar u_R}
\rightarrow \ket{(s_L,\bar u_R);(d_L,\bar d_R)}
\rightarrow \ket{(s_L,\bar d_R);(d_L,\bar u_R)}
\eqno(7)  $$
where we have assumed that helicity is conserved in the gluonic
processes of pair creation and rearrangement into two mesons.
The two mesons are seen to be both produced in helicity zero states
with equal amplitudes for $K^* \pi$ and $K \rho$.
Here the phase space favors $K^* \pi$ in agreement
with experiment. Direct production without final state interactions
give a phase space distribution, rather than a resonance.
If one considers final state interactions as a first approximation via
the $K_1$ (1270) assumed to couple only to $K\rho$ and
the $K_1 (1400)$ assumed to couple only to $K^* \pi$, the two channels
are completely decoupled in the final state rescattering process, and
the dominance of $K^* \pi$ can persist if both final states are enhanced
by the same factor in the resonance region.

Additional information can be obtained by angular distributions and
helicity analysis. In a simple direct-production model with the quark
and antiquark produced at the weak vertex both having ``correct"
helicities and with helicity conservation assumed
in the production of the additional
pair (7), the vector meson is produced in a state of helicity
zero. The angular distribution of its decay in the rest frame of the
vector meson relative to the momentum of the recoiling pseudoscalar
will be $\cos^2\theta$. This angular distribution may be observed with
relatively few events by dividing the angular distribution into
``polar" and ``equatorial" bins separated at $\theta = 60^o$ where
$\cos \theta = 1/2$. A $\cos^2 \theta$ distribution favors the number
of events in the polar bin over those in the equatorial bin by a
factor of seven, whereas an isotropic distribution gives equal numbers
in the two bins.
$$ {{\int_{1/2}^1 \cos^2 \theta d(\cos \theta)}\over
{\int_0^{1/2} \cos^2 \theta d(\cos \theta)}} = 7; ~ ~ ~
{{\int_{1/2}^1  d(\cos \theta)}\over
{\int_0^{1/2} d(\cos \theta)}} = 1
\eqno(8) $$

Such an angular correlation will not occur in the resonance dominance
model. Once the resonance is formed the quark and antiquark are no
longer in helicity eigenstates. The quark antiquark pair is in a state
with $J^P=1^+$ and there is no favored helicity for the decay
vector meson. The polarization of the vector meson in the emitted decay
and the angular distributions depend upon the initial
polarization state of the axial vector resonance, which can be
correlated with the lepton angular distribution, and will also depend
upon the relative amplitudes for $s$ and $d$ waves in the decays of
the resonances. For example, in the case where the decay is pure s wave
the vector meson polarization is identical to the initial polarization
of the axial vector state, all helicity states of the vector meson are
equally populated and the angular distribution of the vector meson
decay momentum relative to the momentum of the recoiling pseudoscalar
is isotropic.

We now consider the possibility that
the weak vertex also produces helicity
suppressed states in which which the quark has the correct helicity and the
antiquark has the wrong helicity and vice versa. The amplitude of such states
is proportional to the mass of the quark or antiquark with the wrong helicity.
Here $GP$ and $G_{us}P$ change a ``correct-helicity-quark, wrong-helicity
antiquark" into a ``correct-helicity-antiquark, wrong-helicity quark" state.
The eigenstates of $GP$ or $G_{us}P$ are therefore linear combinations of
``correct-helicity-quark, wrong-helicity antiquark" and
``correct-helicity-antiquark, wrong-helicity quark" with equal amplitudes and
an
eigenvalue depending upon the relative phase of the two components of the wave
function.

For a quark-antiquark pair with equal masses these two components
in the wave function have equal amplitudes and the state is an eigenstate of
$GP$ in the nonstrange sector and of $G_{us}P$ in the strange sector with a
phase that can be shown to give the negative eigenvalue corresponding to a
first class current and gives a vector state with quantum numbers of the
$\rho$ or $K^*$(890) and no axial vector. For unequal quark masses the two
helicity components in the wave function are not equal and the state is a
linear combination of the two eigenstates of $GP$ or $G_{us}P$ with opposite
eigenvalues. There is thus an additional state which has the positive
eigenvalue of $GP$ or $G_{us}P$ corresponding to a second class current and
the quantum numbers of the $b_1$ or $K_B$.
The treatment as eqs.(2-5) can be extended to the axial vector case, where
the helicity suppressed amplitudes have exactly the same form but with
different phases resulting from the difference between spin-zero and spin-one
states. In the resonance domination model
the relevant states corresponding to the pseudoscalar (first class)
and scalar (second class) states are the $\rho$ or $K^*$(890) (first class)
and the $b_1$ or $K_B$ (second class). The $a_1$ and $K_A$ states which have
the $^3P_1$ configuration in the quark model do not appear here because their
wave functions contain components only with helicity $\pm 1$ and they
do not have a component with zero helicity. This can be
seen by examining the angular momentum coupling of the orbital
angular momentum $L$, total spin $S$ to the total angular momentum $J$.
A helicity zero state must have $L_z=S_z=J_z = 0$ if we choose our z-axis in
the direction of the relative momentum of the quark-antiquark system. Such a
state has no component with the quantum numbers of the $^3P_1$ configuration;
namely $L=1$, $S=1$ and $J=1$, because of the
vanishing of the relevant Clebsch-Gordan coefficient.

In the direct production model the helicity-suppressed state in which
the strange quark has the wrong helicity produces the vector meson in
the helicity one state,
$$
\ket{s_R,\bar u_R}
\rightarrow \ket{(s_R,\bar u_R);(d_L,\bar d_R)}
\rightarrow \ket{(s_R,\bar d_R);(d_L,\bar u_R)}
\rightarrow \ket{\bar K^{*o}\pi^-}
\eqno(9a)  $$
$$
\ket{s_R,\bar u_R}
\rightarrow \ket{(s_R,\bar u_R);(d_R,\bar d_L)}
\rightarrow \ket{(s_R,\bar d_L);(d_R,\bar u_R)}
\rightarrow \ket{\bar K^{o}\rho^-}
\eqno(9b)  $$
These contributions would not be coherent with the dominant contribution
(7)
as they produce a final state with a different helicity. All interference
terms would average out over angular distributions.
Note that since the strong interactions conserve parity, the two
amplitudes (11) are expected to be equal. Thus even when flavor SU(3)
is broken at the weak vertex the direct production model with helicity
conservation predicts equal amplitudes
for $K^* \pi$ and $K \rho$ with phase space favoring $K^* \pi$.

A contribution from the helicity-suppressed transition (11b) wuld show
up in the helicity analysis as a component with
a $\sin^2\theta$ angular distribution for the decay momentum of the
vector meson relative to the momentum of the recoiling pseudoscalar.

The experimental investigation of the strange $\tau$ decays is thus seen
to give information both about weakened helicity suppression and stronger
flavor-symmetry breaking at the pointlike weak vertex arising from
the $u-s$ mass difference, about the hadronization process after the weak
vertex and about the confused hadron spectrum in the
strange scalar and axial vector sectors.
Scalar decays test symmetry-breaking at the weak vertex and the results
are sensitive to the difference between constituent and current quark
masses. Axial decays test strong symmetry breaking and helicity
conservation in the hadronization process, where different models give
different predictions for the mass spectrum, the $K^*\pi/K\rho$ ratio
and helicity amplitudes.
At this stage with statistics showing excesses of only 12 events, it is
difficult to draw conclusions regarding different models. However, when
more data are obtained, a complete partial wave analysis including
helicities may provide useful information.

We can recall that the nonstrange axial vector spectrum and the
existence of the $a_1$ was completely confused until it was clarified by
clean data from $\tau$ decay. Data from strange decays
can provides an even richer source of information since the larger
symmetry breaking allows measurement of symmetry-forbidden processes
which are negligible in the nonstrange sector.

Discussions with M. T. Ronan who called the author's attention to the
experimental results of ref. [\TPC] and to the work of Suzuki (ref [\Suzuki])
are gratefully acknowledged.

\refout
\end